\documentclass[aps,pra,twocolumn,superscriptaddress]{revtex4-2}
\usepackage{epsfig}
\usepackage{natbib}
\usepackage{amsmath,amssymb,mathrsfs}
\usepackage{subfigure}
\usepackage{tabularx}
\usepackage{longtable}
\usepackage{amsfonts}
\usepackage{rotating}
\usepackage{bbold}
\usepackage{hhline}
\usepackage{braket}
\usepackage{txfonts, comment}

\usepackage[colorlinks=true,linkcolor=blue,urlcolor=blue,citecolor=blue]{hyperref}

\begin{document}

\title{Effect of Quantum Statistics on Computational Power of Atomic Quantum Annealers}

\author{Yuchen Luo}
\affiliation{State Key Laboratory of Surface Physics, Institute of Nanoelectronics and Quantum Computing,
and Department of Physics, Fudan University, Shanghai 200433, China}

\author{Xiaopeng Li}
\email{xiaopeng\underline{ }li@fudan.edu.cn}
\affiliation{State Key Laboratory of Surface Physics, Institute of Nanoelectronics and Quantum Computing,
and Department of Physics, Fudan University, Shanghai 200433, China}
\affiliation{Shanghai Qi Zhi Institute, AI Tower, Xuhui District, Shanghai 200232, China}

\date{\today}

\begin{abstract}
Quantum particle statistics fundamentally controls the way particles interact, and plays an essential role in determining the  properties of the system at low temperature. Here we study how the quantum statistics affects the computational power of quantum annealing. We propose an annealing Hamiltonian describing quantum particles moving on a square lattice
and compare the computational performance of the atomic quantum annealers between two statistically-different components: spinless fermions and hard-core bosons. In addition, we take an Ising quantum annealer driven by traditional transverse-field quantum fluctuations as a  baseline. The potential of our quantum annealers to solve combinatorial optimization problems is demonstrated on random 3-regular graph partitioning. We find that the bosonic quantum annealer outperforms the fermionic case. The superior performance of the bosonic quantum annealer is attributed to larger excitation gaps and the consequent  smoother adiabatic transformation of its instantaneous  quantum  ground states. Along our annealing schedule, the bosonic quantum annealer is less affected by the glass order and explores the Hilbert space more efficiently. Our theoretical finding could shed light on constructing atomic quantum annealers using Rydberg atoms in optical lattices. 
\end{abstract}

\maketitle

\section{Introduction}
One of the central challenges in computer science is to design efficient algorithms to solve combinatorial optimization problems \cite{Steiglitz2013combinatorial} that are of practical importance in a broad range of fields. Most of such computational tasks are NP-hard or NP-complete problems, which require minimizing a cost function having a large number of local minima, making it intractable for classical algorithms.

Quantum annealing \cite{Nishimori1998pre, Farhi2001science, Santoro2002science, Das2008rmp, Lidar2018rmp, Hauke2020rpp} is a promising quantum computing approach that may have significant quantum speedup in solving combinatorial optimization problems. It is a heuristic algorithm that formulates the cost function into a quantum many-body Hamiltonian, typically an Ising spin Hamiltonian \cite{Lucas2014fip}, then utilizes quantum fluctuations to escape from trapping by local minima and explores the spin-glass-like energy landscape for the ground state that encodes the optimal solution. The theoretical foundation of quantum annealing is built on the adiabatic theorem of quantum mechanics \cite{Kato1950jpsj, Messiah1962quantum}, which ensures that the quantum system will stay close to the instantaneous ground state if the initial Hamiltonian changes into the problem Hamiltonian slowly enough, and the annealing time scales inversely proportional to a polynomial of the minimal energy gap \cite{Sabine2007jmp, Nishimori2008jmp, Lidar2009jmp}.

Holding the expectation of quantum speedups over algorithms running on classical computers, numerous efforts have been devoted to realization of large-scale programmable quantum devices, from superconducting qubits to trapped ions \cite{Hauke2020rpp, Altman2021prx}. There have been rapid technological advances in recent years in fabricating quantum annealers having a large number of spins, which has been attracting extensive attention in industry and academia \cite{Johnson2011nature, Lidar2013nc, Sergio2014nphys}. Nevertheless, for technical issues such as quantum decoherence and control errors, whether quantum annealing could fulfill its promise of significant quantum speedup over classical computing is still an open question. In the NISQ era \cite{Preskill2018quantum}, developing alternative quantum annealing protocols and architectures with near-term technology is still in great demand for reaching the quantum computation advantage on combinatorial optimization problems.

With remarkable experimental progress in the last two decades, ultracold atoms in optical lattices, originally engineered as a highly-controllable quantum simulator for exotic many-body physics, provide new opportunities to build a scalable quantum annealer \cite{Bloch2017science, Takasu2020nrp}. In addition to refined measurement techniques, free programmability in optical lattices has been recently achieved via single-site control and cavity-mediated long-range interactions \cite{Gauthier2016opt, McDonald2019prx, Subhankar2019prx, Mazurenko2017nature, Yi2008iop, Qiu2020npj, Torggler2017pra, Torggler2019quantum, Alexey2021prx}, generically necessary to carry out classical optimization on a quantum annealer. Based on the atomic platform, it is a natural choice to encode a qubit into the occupation number of a lattice site, which can be either zero or one of intrinsically-repulsive spinless fermions or hard-core bosons. The atomic tunnelings, in which the effect of quantum  statistics is embodied, play the role of quantum fluctuations to drive the search for the optimal occupation configurations. Quantum particle statistics, as a fundamental concept in quantum physics, is an important factor in determining the dynamical properties of the system, and can give rise to big contrasts in the nature of the ground state \cite{Kuklov2003prl, Rigol2010pre, Rigol2012pra, Nie2018prb}. Following are the questions that which of the fermionic or the bosonic tunnelings, serving as quantum fluctuations, leads to better computational performance, and how the two atomic quantum annealers with different quantum  statistics behave differently during annealing. It is worthy pursuing this direction from both theoretical interests and practical considerations. 

Here we propose an atomic quantum annealer under fermionic or bosonic statistics and investigate the effect of quantum statistics in this context. A commonly used model for quantum annealing is the Ising spin Hamiltonian which is universal for classical problems \cite{Cubitt2016science} and has been implemented in a range of different physical platforms \cite{Johnson2011nature, Kim2011iop, Bernien2017nature}. For this reason, we also compare the atomic quantum annealer with the Ising-model approach. To carry out a concrete analysis, we focus on the performance of the three quantum annealers to solve the random instances of 3-regular graph partitioning using a fixed amount of annealing time. By numerical simulation of the time evolution of our quantum annealers, we find that, on average, the bosonic quantum annealer reaches higher success probability than the fermionic one, and the performance of the bosonic quantum annealer is comparable with the Ising-model quantum annealer. How the three quantum annealers perform is directly related to their low-energy spectra, and the bottlenecks can be diagnosed by the ground-state fidelity susceptibility which measures the transformation smoothness of the instantaneous quantum ground  state. Looking into their behavior during annealing, the bosonic quantum annealer is always less affected by the glass order and explores a larger portion of the many-body Hilbert space, as compared to the fermionic one.  

The paper is organized as follows. In Sec.~\ref{sec:2}, we first introduce the graph partitioning problem and describe the setup of our atomic quantum annealers. In Sec.~\ref{sec:3}, we numerically make a comparison among the computational performance of the three quantum annealers and study several properties to illustrate their behavior difference. Finally, our discussion and conclusions are presented in Sec.~\ref{sec:4}.

\section{Quantum Annealing Setup}\label{sec:2}

\subsection{Graph partitioning}
Partitioning is one of the six basic NP-complete problems proposed by Garey and Johnson in 1979 \cite{garey1979computers}. This class of problems bears close resemblance to the spin glass problem. Studying the complexity hardness of partitioning problem and understanding the physics behind spin glasses are auxiliary to each other \cite{Barahona1982iop, Fu1986iop}. 

Here we focus on the well-studied problem graph partitioning \cite{Lucas2014fip}, which is defined on an undirected graph $G=(V,E)$, with an even number $N=|V|$ of vertices connecting by a set of edges $E=\{(v_i,v_j)\}$. The task is to partition the vertex set $V$ into two subsets $V_1$ and $V_2$ with equal size $N/2$ such that the number of edges connecting $V_1$ and $V_2$ is minimized. Then the cost function can be formulated as this edge number
\begin{equation}
   C(V_1,V_2) = \sum_{v_i\in V_1,v_j\in V_2} A_{ij}, 
   \label{eq:costfunction}
\end{equation}
where $A_{ij}$ is the adjacent matrix element of the graph, equal to one when there is an edge between vertex $v_i$ and vertex $v_j$, and zero when there is no edge. We consider random graph instances in a specific ensemble where the degree of every vertex is fixed to three, and by definition each row and column of the adjacent matrix $A$ sums to this fixed degree.

\subsection{The atomic quantum annealer}

Considering that spinless fermions or hard-core bosons can only occupy a lattice site by particle number zero or one, we naturally encode a qubit into a site occupation and the solutions of combinatorial optimization problems into the occupation configurations. For graph partitioning, each vertex $v_i$ on the graph corresponds to a lattice site $i$ and the graph connectivity is embedded in the interactions between atoms. In this way, the problem Hamiltonian in the atomic quantum annealer is of the form
\begin{equation}
   H_{\text{P}}^{\text{atomic}} = \sum_{(v_i,v_j)\in E} \frac{1}{2}\left[1-\left(2n_i-1\right)\left(2n_j-1\right)\right], 
   \label{eq:HPBF}
\end{equation}
where the site occupation $n_i=\{0,1\}$ denotes the corresponding vertex being in either of two subsets. The non-local graph connectivity can be realized by adopting the quantum wiring scheme~\cite{Qiu2020prx}, which has been realized by controlling Rydberg excitations in an atomic experiment~\cite{2022_Ahn_NatPhys}. Since the total particle number is conserved in an atomic system, the equal-partitioning constrain is satisfied by taking the particle number to be half-filling $N_p=\sum_{i=1}^N n_i =N/2$.

The nearest-neighbor atomic tunneling on the two-dimensional lattice
\begin{equation}
   H_{\text{T}} = \sum_{\langle i,j\rangle} -\left( a_i^{\dagger}a_j+a_j^{\dagger}a_i\right),
   \label{eq:HT}
\end{equation}
serves as the driver Hamiltonian to provide quantum fluctuations of occupation configurations, where $\langle i,j\rangle$ labels a pair of nearest-neighbor sites and $a_i^{\dagger}$($a_i$) creates (annihilates) a spinless fermion or a hard-core boson at site $i=(m,n)$. Open boundary condition is considered here.

To initialize the quantum annealer with an easily-implemented ground state that is unique and finitely-gapped, we choose the initial Hamiltonian to be a set of onsite potentials
\begin{equation}
   H_{\text{V}} = \sum_{i=1}^{N} V_i n_i, 
   \label{eq:HV}
\end{equation}
which takes a specific occupation configuration $|0101...\rangle$ as its ground state by setting $V_i=-2$ for even $i$ and $V_i=0$ for odd $i$. The bias of this artificial choice is expected to be eliminated by random graph sampling.

We propose an annealing schedule that linearly interpolates from $H_{\text{V}}$ to $H_{\text{P}}^{\text{atomic}}$ with 
\begin{equation}
   H^{\text{atomic}}(s) = \left(1-s\right)H_{\text{V}} + \lambda s \left(1-s\right) H_{\text{T}}+ s H_{\text{P}}^{\text{atomic}}. 
   \label{eq:HtimeBF}
\end{equation} 
Here, the Hamiltonian $H_\text{T}$ is introduced to generate quantum fluctuations. This is necessary in our scheme because the initial and final Hamiltonians commute, i.e., $[H_{\text{V}}, H_{\text{P}}^{\text{atomic}}] =0$. 
The parameter $s(t)=t/\mathcal{T} \in [0,1]$ determines the annealing path with the total annealing time $\mathcal{T}$, and the parameter $\lambda$, set to be $\lambda=3$, controls the driving strength. 

We consider two atomic quantum annealers that are respectively assembled with spinless fermions and hard-core bosons in order to investigate the effect of quantum statistics on the annealing performance. Although their annealing processes are described by a Hamiltonian of the same form (\ref{eq:HtimeBF}), the two annealers are fundamentally different, for the underlying quantum particles obey different statistics. Their quantum dynamics are drastically different.  
The bosonic tunneling is equivalent, via Jordan-Wigner transformation \cite{Jordan1928physik}, to the XXZ model defined on the same 2D lattice
\begin{equation}
   H_{\text{XXZ}} = \sum_{\langle i,j\rangle} -\left(\sigma_i^x\sigma_j^x+\sigma_i^y\sigma_j^y\right),
\end{equation}
whose dynamics conserved the total magnetization $S_z = \sum_{i=1}^N \sigma_i^z$, i.e., $[H_{\text{XXZ}},S_z]=0$. The XXZ model based quantum annealing has been suggested for performing constrained optimization \cite{Hen2016pra}. In the fermionic case, due to Fermi-Dirac quantum statistics, the Jordan-Wigner transformation of the tunneling Hamiltonian would generate non-local Jordan-Wigner strings, producing highly nontrivial sign structures for the Hamiltonian matrix in the computational basis. It is not a priori clear whether the non-local Fermi sign structure would be helpful or harmful to the quantum annealing \cite{Hormozi2017prb,Albash2019pra,Takada2020jpsj,Crosson2020Quantum,Takada2021prr}.
On one hand, it generates non-local couplings, which would tend to make the many-body system more ergodic and potentially weaken the spin glass problem \cite{Abanin2019rmp}; on the other hand, it involves non-stoquastic Hamiltonians at intermediate times whose ground states are in general difficult to reach \cite{Lubasch2011prl, Chiu2018prl, Lin2019prl}.

\subsection{The Ising quantum annealer}
Encoded as an Ising model, the problem Hamiltonian for graph partitioning consists of two parts
\begin{equation}
   H_{\text{P}}^{\text{Ising}} = \sum_{(v_i,v_j)\in E} \frac{1}{2}\left(1-\sigma_i^z \sigma_j^z\right) + \alpha \left(\sum_{i=1}^{N} \sigma_i^z\right)^2, 
   \label{eq:IsingHp} 
\end{equation}
where the first term carries the energy cost of edges connecting vertices from different subsets, and the second provides an energy penalty representing the size imbalance of the two subsets. In order to ensure the balancing constraint in the equal-partition problem, the factor $\alpha$ for graph partitioning must satisfy the condition $\alpha\ge \frac{1}{8} \text{min} (2\Delta,N)$, with $\Delta$ the maximal degree of graph~\cite{Lucas2014fip}. We choose a fairly small value for $\alpha = 1$ that satisfies this condition, as a too large penalty would suppress quantum fluctuations in the adiabatic quantum evolution and make the quantum annealing inefficient.  

Since we focus on analyzing the effect of quantum statistics on the quantum annealing in this study, we take a Hamiltonian schedule for the Ising model based quantum annealing, 
\begin{equation}
   H^{\text{Ising}}(s) = \left(1-s\right)H_{\text{Z}} + \lambda s \left(1-s\right) H_{\text{X}}+ s H_{\text{P}}^{\text{Ising}}. 
\end{equation}
This Hamiltonian schedule is chosen to be similar to the atomic quantum annealer in Eq.~\eqref{eq:HtimeBF} for a fair comparison. The Hamiltonian schedule starts from the longitudinal-field Hamiltonian
\begin{equation}
   H_{\text{Z}} = \sum_{i=1}^N h_i\sigma_i^z, 
\end{equation}
where onsite fields have equal amplitude but alternating signs --- $h_i=-1$ for even $i$ and $h_i=1$ for odd $i$, in correspondence with the initial potential (\ref{eq:HV}) of the atomic quantum annealer. 
At intermediate time with $0<s<1$, we have introduced the transverse-field Hamiltonian,  
$H_{\text{X}} = -\sum_{i=1}^N \sigma_i^x$,
to drive quantum fluctuations. This Ising quantum annealer offers a performance reference point to benchmark the fermionic and bosonic atomic quantum annealers.

\section{Numerical Results}\label{sec:3}

\subsection{Success probability}
The computation performance is characterized by the success probability of the quantum annealer, namely the total probability of finding the correct solution in the projective measurement of the final quantum state $|\psi_{\text{f}}\rangle$, 
\begin{equation}
   P_{\text{s}}(\mathcal{T}) = \sum_{i=1}^D |\langle \psi^i_{\text{s}}|\psi_{\text{f}}\rangle|^2, 
   \label{eq:Ps}
\end{equation}
with $|\psi ^i _{\text{s}}\rangle$ the ground states of the final Hamiltonian, $i=1,2, \ldots, D$ labeling the ground state degeneracy.  
We study three problem sizes with the 2D lattice geometries $N = 4\times 3, 4\times 4, 4\times 5$, and generate $1000$ random problem instances for each system size. The total annealing time is fixed to $\mathcal{T}=50$, and for each instance we calculate the success probabilities of our fermionic, bosonic and Ising quantum annealers.

\begin{figure}[htp]
   \centering
   \includegraphics[width=0.48\textwidth]{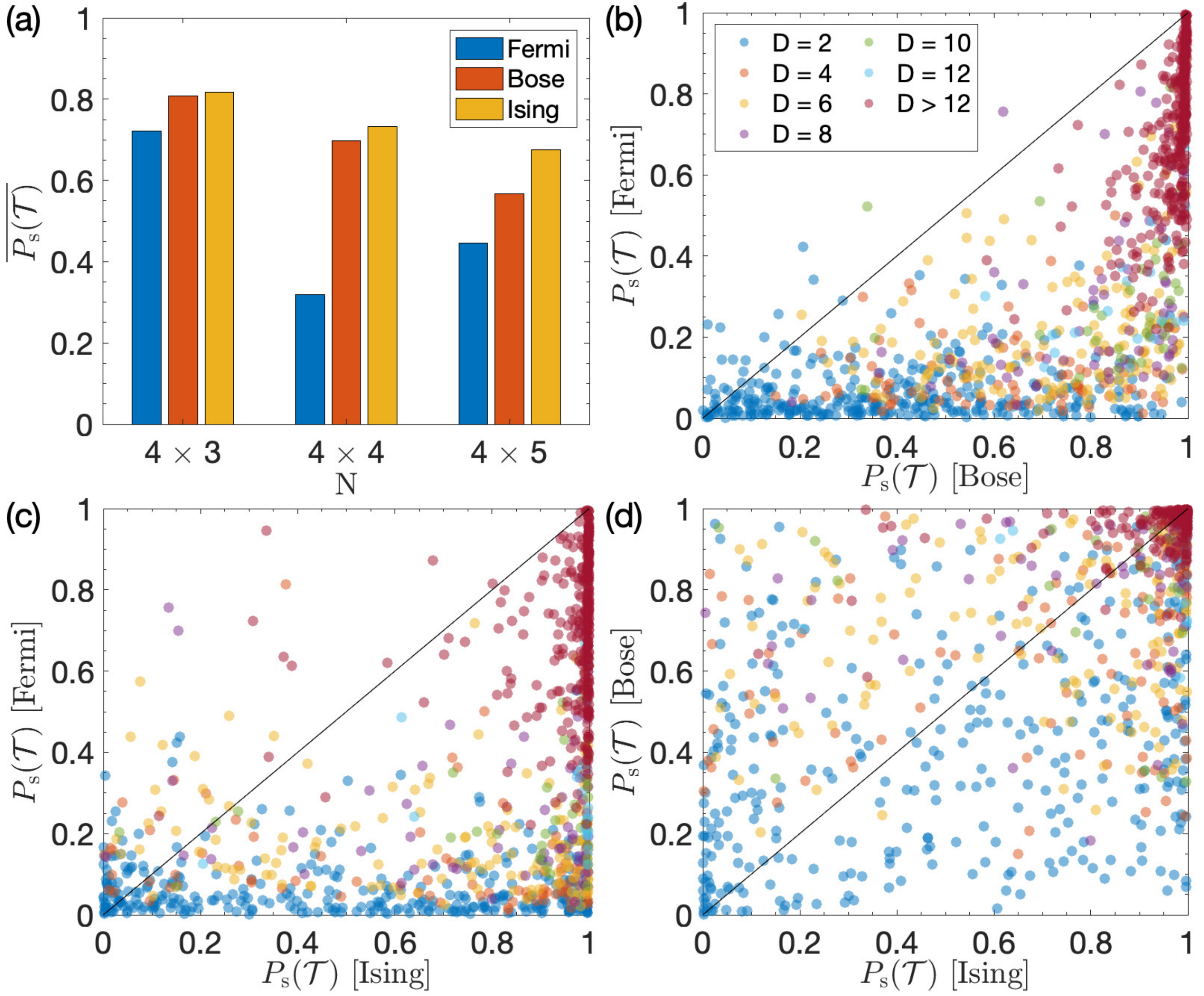}
   \caption{The success probabilities of our fermionic, bosonic and Ising quantum annealers with the total annealing time $\mathcal{T}=50$. (a) The instance-averaged success probabilities of the three quantum annealers for the problem sizes $N = 4\times 3, 4\times 4, 4\times 5$. The individual success probabilities for the problem size $N = 4\times 4$, compared between (b) the fermionic and the bosonic quantum annealers, (c) the fermionic and the Ising quantum annealers, and (d) the bosonic and the Ising quantum annealers. Each dot represents one instance, and its color denotes the solution degeneracy $D$ of this instance.}
   \label{fig:1}
\end{figure}

The performance comparison among the three quantum annealers is presented in Fig.~\ref{fig:1}. As shown in Fig.~\ref{fig:1} (a), the fermionic quantum annealer has the worst performance for all problem sizes compared to bosonic and Ising quantum annealers. The performance of the bosonic and the Ising quantum annealers is more or less comparable. Focusing on the problem size $N=4\times 4$, we then look into the success probability for each instance, and compare the success probability distributions of the three quantum annealers. The pairwise comparisons of the individual success probabilities are shown in Fig. \ref{fig:1} (b-d), and the instances with different solution degeneracy are represented by the dots with different colors. For almost all instances ($97.3\%$), the success probabilities of the bosonic quantum annealer are higher than that of the fermionic one. Comparing the two atomic quantum annealers with the Ising quantum annealer, the fermionic annealer produces higher success probability than the Ising annealer only for very rare instances with a small percentage of $0.095\%$,  
whereas the rate of the bosonic quantum annealer outperforming the Ising annealer is significant reaching $36.4\%$.
There are a number of problem instances that the bosonic quantum annealer way outperforms the Ising annealer.

\begin{figure}[htp]
   \centering
   \includegraphics[width=0.48\textwidth]{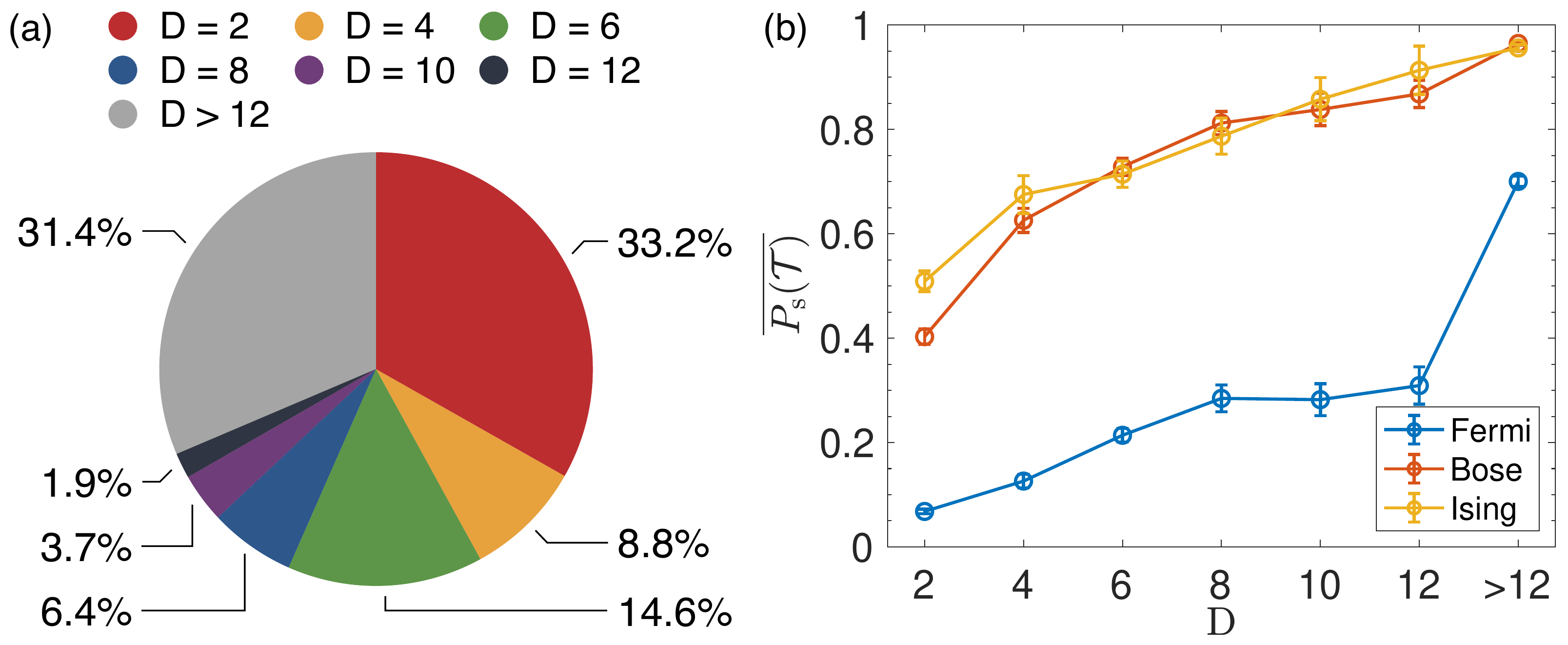}
   \caption{(a) The solution degeneracy distribution of our 1000 problem instances for the system size $N=4\times 4$. (b) The success probabilities of our fermionic, bosonic and Ising quantum annealers for the problem size $N=4\times 4$, which are averaged over the instances with the same solution degeneracy $D$ and presented as functions of $D$.}
   \label{fig:2}
\end{figure}

For our three quantum annealers, the individual success probabilities exhibit consistent correlations with the solution degeneracy $D$. As a result of the bit-flip symmetry of graph partitioning
\[
\sigma_i ^z \to -\sigma_i ^z, 
\] 
the solution degeneracy $D$ only takes even numbers, and the two final ground states up to a bit-flip produce two equivalent partition solutions. The solution degeneracy $D$ of our generated instances for the problem size $N=4\times 4$ has a broad distribution from $D=2$ up to $D=96$. The distribution of $D$ is shown in Fig. \ref{fig:2} (a). The problem instances with $D=2$, i.e., having one unique partition solution, count as $1/3$ of all instances. We then divide all problem instances into different classes according to the degeneracy $D$, and average over the quantum annealing success probability within each class. The results are shown in Fig.~\ref{fig:2} (b).  
We find consistent correlations between the success probability and the solution degeneracy for the average success probabilities of all three quantum annealers, in all of which, the success probability grows monotonically with $D$.  This implies that solving partitioning problems by quantum annealing is easier as the number of its solutions gets larger. 
The fermionic quantum annealer has much lower average success probabilities than the bosonic and the Ising ones for all classes of different solution degeneracies. The overall performance of the bosonic quantum annealer is comparable to the Ising case. 

The above comparison of success probabilities indicates that the bosonic tunneling, that drives the quantum fluctuations at intermediate time, possesses stronger computational power than the fermionic tunneling. The computation performance of the bosonic quantum annealer is comparable with the Ising model. One common feature between bosonic and Ising quantum annealers is that their Hamiltonians in the computational basis are both stoquastic \cite{bravyi2006arxiv}. This in sharp contrasts with the fermionic quantum annealer where the Hamiltonian is non-stoquastic.

\subsection{Relevant gap and low-energy property}
From the quantum adiabatic theorem \cite{Kato1950jpsj, Messiah1962quantum}, the success probability of quantum annealing is governed by the minimum gap between the instantaneous ground state and the first excited state. In our situation where the ground state of the problem Hamiltonian is degenerate, it is reasonable to define a relevant gap  between the instantaneous ground state and the first excited state outside the degenerate ground state subspace, 
\begin{equation}
   \Delta E_{\text{R}} = \text{min}_{s\in[0,1]} [E_D(s)-E_0(s)],
\end{equation} 
which sets an upper bound for the minimum gap between the ground state and the first excited state of the instantaneous Hamiltonian during the whole annealing process.

\begin{figure}[htp]
   \centering
   \includegraphics[width=0.48\textwidth]{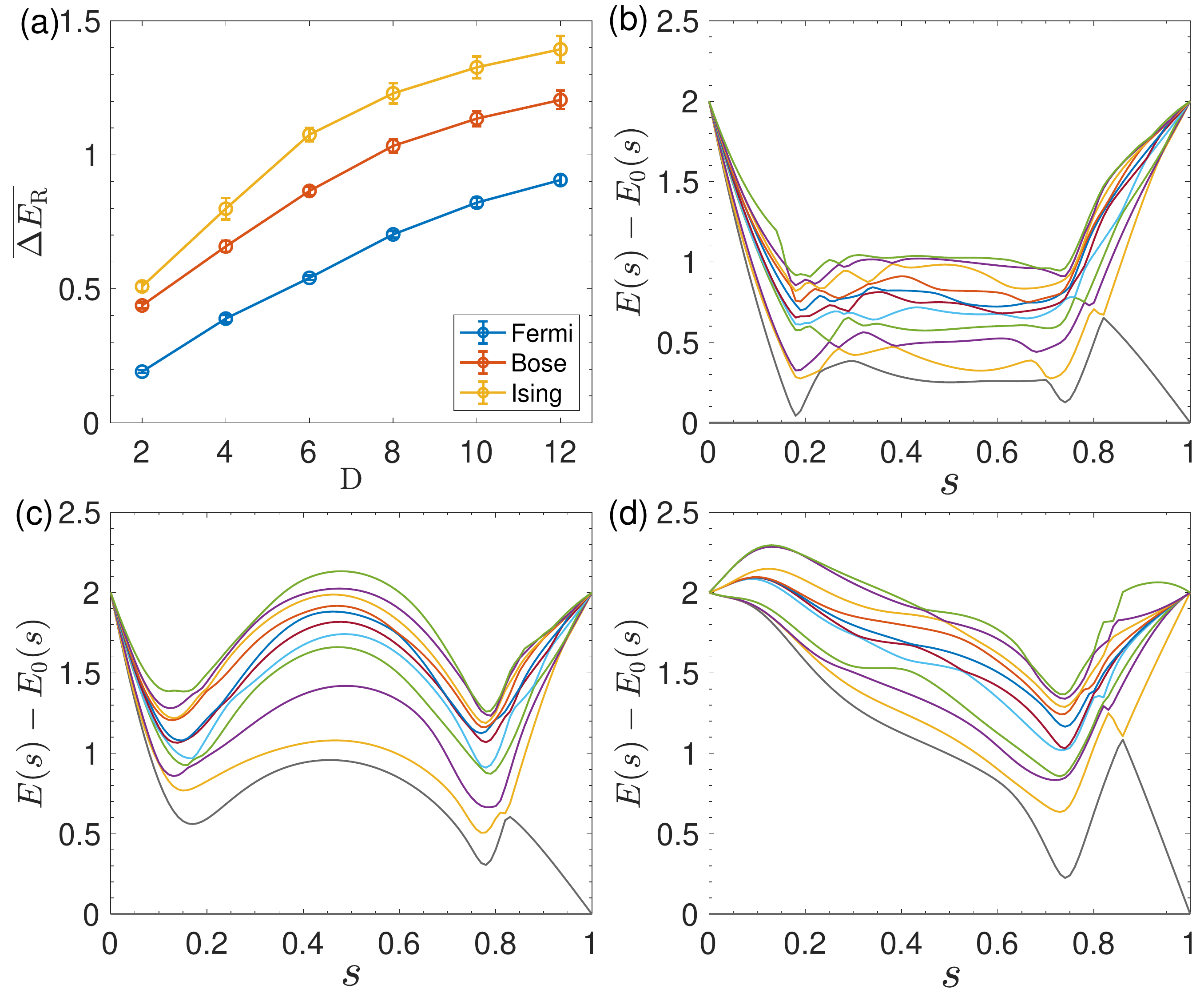}
   \caption{(a) The relevant gaps of the interpolating Hamiltonians followed by the fermionic, the bosonic and the Ising quantum annealer to solve the problem with size $N=4\times 4$, which are averaged over the instances with the same solution degeneracy $D$ and presented as functions of $D$. The lowest twelve energy levels of the interpolating Hamiltonians to solve a typical instance with $N=4\times 4$ and $D=2$ by (b) the fermionic, (c) the bosonic and (d) the Ising quantum annealers. The lowest two grey lines represents the energy levels converging to the two-degenerate ground states of the problem Hamiltonian, and the other colorful lines corresponds to excited states.}
   \label{fig:3} 
\end{figure}

Along our annealing schedule to solve the problem with size $N=4\times 4$, the relevant gaps are numerically determined for each instance, and averaged over the instances with the same solution degeneracy $D$. The results for our three quantum annealers are presented in Fig.~\ref{fig:3} (a) as functions of $D$. We observe systematic increase in the averaged relevant gap with the solution degeneracy, which is consistent with the behavior of success probability as shown in Fig.~\ref{fig:2} (b). 
The fermionic quantum annealer is found to have the smallest relevant gaps for all the solution degeneracies, which we expect to be the main reason for its computation performance being the worst as compared to the bosonic and the Ising models. Consequently, the required evolution time for the fermionic quantum annealer to reach a certain success probability threshold is expected to be significantly larger compared to the other two cases. 

We further concentrate on the problem instances with $D=2$, the largest proportion and the hardest to solve, and look into their low-energy properties relating to the performance difference. Since the performance of quantum annealing can be directly reflected by the low-energy spectrum, we randomly choose a problem instance and show the typical structure of the low-energy levels of the interpolating Hamiltonians to solve this instance by the three quantum annealers in Fig. \ref{fig:3} (b-d).
For our two atomic quantum annealers, there are two minimum gaps located at the two stages where quantum fluctuations are turning on and off. The minimum gaps of the fermionic quantum annealer are much smaller than the bosonic annealer.  
In addition, the low-energy spectrum of the fermionic quantum annealer, compared to that of the bosonic one, has a denser structure and involves more level crossings during the annealing process. The Ising quantum annealer has very different low-energy spectrum from the atomic ones. It develops only one minimum gap at the late stage where quantum fluctuations are turning off.

Gap closing during the annealing process is generally induced by quantum phase transition \cite{Krzakala2008prl, Amin2009pra, Krzakala2010prl, Farhi2012pra}. Among various probes of quantum phase transition, the ground-state fidelity susceptibility per site is a universal indicator, regardless of the transition mechanism, to locate the quantum critical point and the attendant minimum gap \cite{Gu2010IJMPB}. This quantity is defined as the second derivative of the fidelity $\mathcal{F}(s,s+\delta s)=|\langle \psi_{\text{g}}(s)|\psi_{\text{g}}(s+\delta s)\rangle |$ to measure the degree of criticality by the rapidity of the ground-state variation as a function of the system parameters:
\begin{equation}
   \mathcal{S}(s) = \lim_{\delta s\to 0} \frac{-2\rm{ln}\mathcal{F}(s+\delta s)}{N(\delta s)^2}.
\end{equation}
A peak in this fidelity susceptibility signals approaching certain quantum critical point where the ground-state wave function changes dramatically. It is more difficult to maintain quantum adiabaticity at high $\mathcal{S}(s)$.

\begin{figure}[htp]
   \centering
   \includegraphics[width=0.4\textwidth]{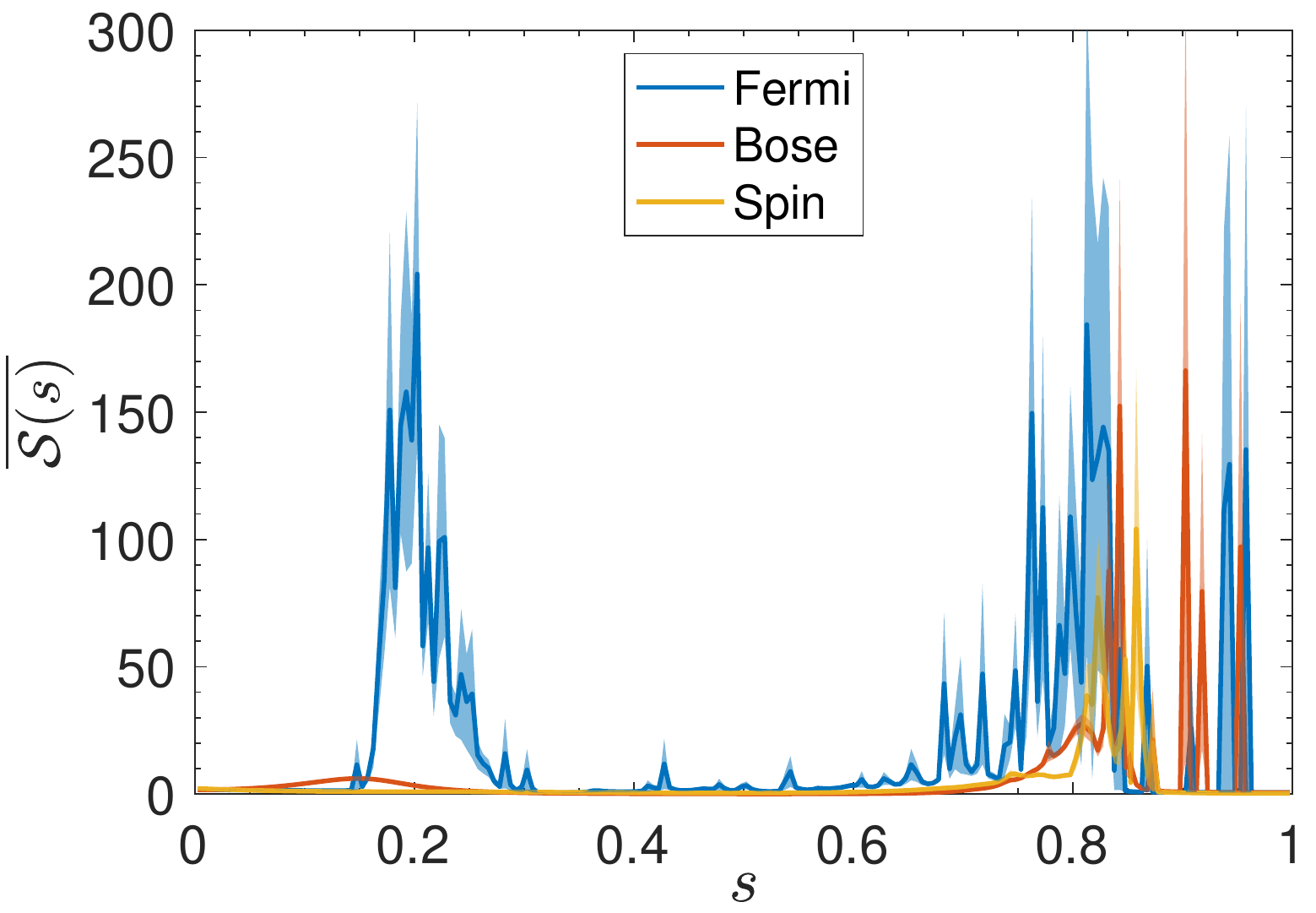}
   \caption{The ground-state fidelity susceptibilities per site along the annealing process of the fermionic, the bosonic and the Ising quantum annealers, which are averaged over the instances with $N=4\times 4$ and $D=2$.}
   \label{fig:4}
\end{figure}

The average ground-state fidelity susceptibility per site along the annealing process to solve the instances with $N=4\times 4$ and $D=2$ by our three quantum annealers is shown in Fig. \ref{fig:4}. For the atomic quantum annealers, two obvious peaks in the fidelity susceptibility occur, respectively corresponding to the two minimum gaps at the two stages with quantum fluctuations turning on and off. The peaks for bosonic quantum annealer, especially the first one, are less prominent than the fermionic one, which indicates the ground state of the bosonic quantum annealer transforms more smoothly along the annealing process. As expected, the fidelity susceptibility of the Ising quantum annealer rises up significantly at the appearance of its minimum gap, and the peak value is comparable to that of the bosonic annealer at the same location.

Although a general framework describing the quantum annealing efficiency is absent, it is widely believed that quantum annealing might be bottlenecked by the glass phase \cite{Knysh2016nc}. The glass phase appears below some critical value of quantum fluctuations, and is characterized by energy spectrum near ground states. In this phase, small changes in Hamiltonian parameters may lead to a chaotic reordering of associated energy levels, which causes level crossings with exponentially small energy gaps. In order to examine the glass physics in the quantum annealers, we calculate the Edwards-Anderson order parameters of the glass phase~\cite{Edwards1975iop}. 
For the atomic quantum annealers, we measure the fluctuations of occupation numbers on each lattice site, 

\begin{equation}
   q_n= \frac{1}{N}\sum_{i=1}^N\langle 2n_i-1\rangle ^2. 
\end{equation}
For the Ising quantum annealer, we measure the fluctuations of spin orientation, 
\begin{equation}
   q_z= \frac{1}{N}\sum_{i=1}^N\langle \sigma_i^z\rangle ^2. 
\end{equation}
These two expressions are both normalized to the range of $[0,1]$. 

\begin{figure}[htp]
   \centering
   \includegraphics[width=0.38\textwidth]{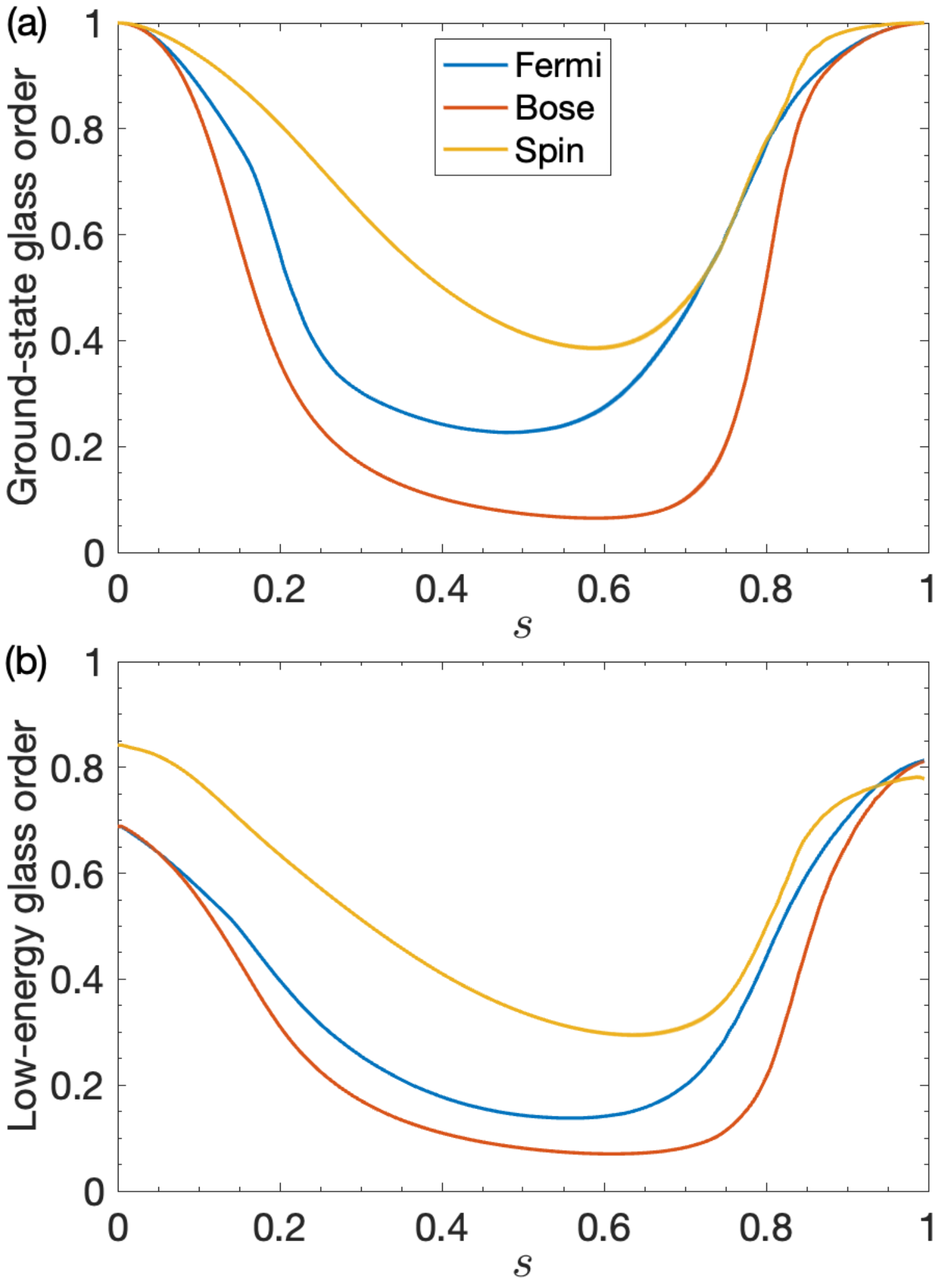}
   \caption{(a) The ground-state glass order and (b) the average glass order of the lowest twelve eigenstates along the annealing process of the fermionic, the bosonic and the Ising quantum annealers. Both are averaged over the instances with $N=4\times 4$ and $D=2$.}
   \label{fig:5}
\end{figure}

The glass order strength along the annealing process of the three quantum annealers is shown in Fig. \ref{fig:5}, which is averaged over different problem instances with $N=4\times 4$ and $D=2$. We calculate the ground-state glass order, as well as the low-energy glass order that is averaged over the lowest twelve eigenstates of the interpolating Hamiltonian.
The two quantities involving the low-energy states are largely related to the performance of quantum annealers. Along the whole annealing process, from an initial product state to the final ground state, the bosonic quantum annealer is consistently less affected by the glass order compared to the fermionic one. This is consistent with our observation on the low-energy spectra in Fig. \ref{fig:3} (b,c), where the bosonic quantum annealer develops less anti-crossings. Although the comparable computation performance with the bosonic quantum annealer, the Ising quantum annealer always stays in stronger glass order, indicating its instantaneous states are very different from the bosonic one.

\subsection{Annealing dynamical behavior} 
We further investigate difference of the three quantum annealers in the dynamical behavior. The problem instances with $N = 4\times 4$, and $D=2$ are still taken for illustration. We calculate the effective dimension of the dynamical state, which characterizes how efficiently the intermediate dynamical state explores the entire Hilbert space. It is defined as, 
\begin{equation}
   D_{\text{eff}}(|\psi\rangle)= \left(\sum_{i=1}^{\mathcal{D}}|c_i|^4\right)^{-1},
\end{equation}
where the $c_i$s are the amplitudes of the state $|\psi\rangle$ on each computational basis, and $\mathcal{D}$ is the dimension of the Hilbert space. This quantity has been widely used in the study of  delocalization or thermalization \cite{Linden2009pre, Short2012iop, Iyoda2018prd}.

\begin{figure}[htp]
   \centering
   \includegraphics[width=0.42\textwidth]{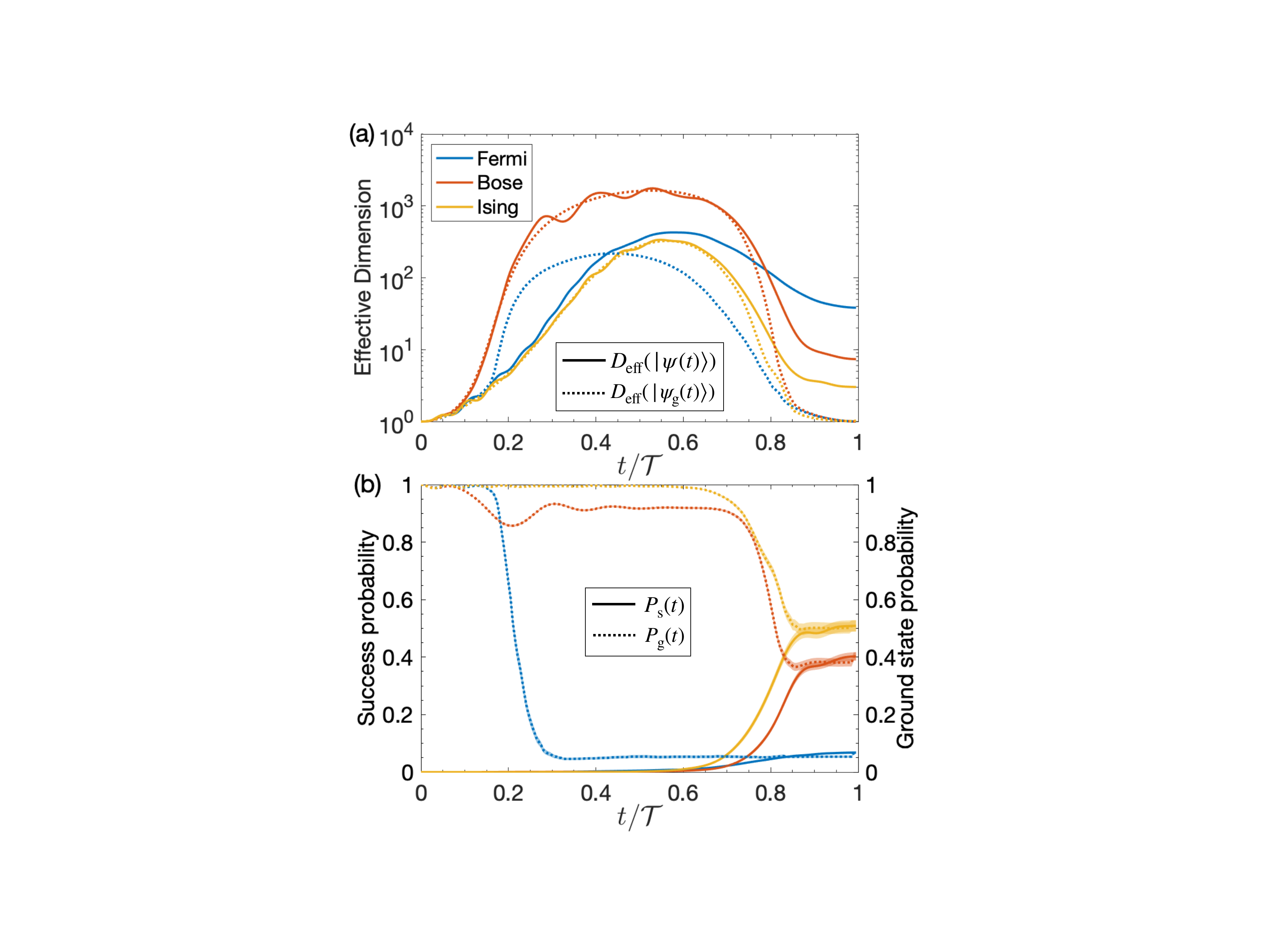}
   \caption{(a) The effective dimensions of the time-evolving state $|\psi(t)\rangle$ (solid lines), with the ones of the instantaneous ground state $|\psi_{\text{g}}(t)\rangle$ (dotted lines) as an adiabatic reference, and (b) the instantaneous values of the success probability (solid lines) and the ground state probability (dotted lines) along the annealing process of the fermionic, the bosonic and the Ising quantum annealers. The above results are averaged over the instances with $N=4\times 4$ and $D=2$.}
   \label{fig:6}
\end{figure}

We consider the effective dimensions of the time-evolving state $|\psi(t)\rangle$ along the annealing process for each quantum annealer, with the ones of the instantaneous ground state $|\psi_{\text{g}}(t)\rangle$ as an adiabatic reference. The results are shown in Fig. \ref{fig:6} (a). All of the three quantum annealers drive their initial states to a large fraction of the Hilbert space as quantum fluctuation turns on. The quantum dynamical state gradually converges to the solution configurations with the ramp-off of the quantum fluctuation. The bosonic quantum annealer expands its dynamical state to the large Hilbert space with a faster rate than the fermionic annealer. The resultant peak value of the effective dimension at the intermediate time is also significantly larger for the bosonic quantum annealer than the fermionic case. Although the value of effective dimensions of Ising quantum annealer is significantly smaller than the bosonic annealer, their qualitative behavior is similar. For both of them, with the ramp-up of quantum fluctuations, the dynamical quantum state follows closely with the instantaneous ground state, and the increase of the effective dimensions in the dynamical state agrees with the instantaneous ground state. This is in sharp contrast with the fermionic quantum annealer where the effective dimension of the dynamical state deviates significantly from the instantaneous ground state.

Furthermore, we look into the instantaneous values of the success probability and the ground state probability along the annealing process of our three quantum annealers. The success probability $P_{\text{s}}(t)$ follows the definition of Eq. \ref{eq:Ps}, and the instantaneous ground state probability $P_{\text{g}}(t)$ is defined as the probability of the time-evolving state $|\psi(t)\rangle$ populating on the instantaneous ground state $|\psi_{\text{g}}(t)\rangle$,
\begin{equation}
   P_{\text{g}}(t) = |\langle \psi_{\text{g}}(t)|\psi(t)\rangle|^2.
   \label{eq:Pg}
\end{equation}
As shown in Fig. \ref{fig:6} (b), the success probabilities $P_{\text{s}}(t)$ of all three quantum annealers stay almost vanishing for about half of the annealing process, and then increase at different rates to their final success probabilities $P_{\text{s}}(\mathcal{T})$ as the annealing Hamiltonian approaches to the problem Hamiltonian. Regarding the ground state probability $P_{\text{g}}(t)$, the three quantum annealers exhibit distinct behaviors within our annealing time. 
The value of the bosonic goes through two obvious drops --- one relatively smooth drop near the beginning and the other a steeper drop near the final. 
The fermionic quantum annealer develops an abrupt drop at the early stage and its ground state probability remains small until the end.
As for the Ising quantum annealer, its ground state probability $P_{\text{g}}(t)$ does not drop until a late time when quantum fluctuations are turning off. The dynamical behaviors of the ground state probability $P_{\text{g}}(t)$ of the three quantum annealers are consistent with our observations on their minimum gaps in Fig. \ref{fig:3} (b-d) and the deviations of their effective dimensions in Fig. \ref{fig:6} (a).

\section{Discussion and Conclusion}
\label{sec:4}
In this work, we have carried out a systematic analysis on the effect of quantum statistics on the computational power of atomic quantum annealers. We propose an annealing schedule for our atomic quantum annealers, and the effect of quantum statistics is embodied in the driver Hamiltonian, i.e.,the fermionic or the bosonic tunnelings. In addition, an Ising quantum annealer is considered to provide a performance reference. The performance comparison among the three quantum annealers is demonstrated by  solving random problem instances of 3-regular graph partitioning in a fixed annealing time. 
For all the problem sizes considered, the numerical results of their final success probabilities show that the bosonic quantum annealer outperforms the fermionic one, reaching comparable performance with Ising quantum annealer.

To shed light on the difference of the three quantum annealers, we study the problem instances with $N=4\times 4$ as an illustration, and further concentrate on the subset with $D=2$ for more details. The computational performance of the three quantum annealers is largely determined by their minimum gaps, which are reflected by their low-energy spectra. For two atomic quantum annealers, two minimum gaps emerge respectively when the quantum fluctuations are turned on and off.  
The first minimum gap of the fermionic quantum annealer is much smaller than that of the bosonic annealer, which is the major bottleneck of its computational performance. The Ising quantum annealer has only one minimum gap at the late stage when quantum fluctuations are turning off. The emergence of their minimum gaps is consistent with the peaks in their ground-state fidelity susceptibility which measures the smoothness of the state transformation during the annealing. Along our annealing schedule, the bosonic quantum annealer suffers from weaker glass order and explores the Hilbert space more extensively than the fermionic one. 

From the consideration of experimental realization, the atomic quantum annealer is particularly suitable for optimization problems with the constraint of the form $\sum_{i=1}^N \sigma_i^z=c$, which is automatically satisfied under particle conservation. For the traditional Ising quantum annealer, it is a standard method to impose constraints by adding penalty terms \cite{Lucas2014fip}. The penalty terms generally requires for all-to-all connectivity that brings enormous challenge to the near-term quantum device. The atomic quantum annealer is free of all complications resulting from penalty terms and cuts down the resource to encode a problem dramatically.

\begin{acknowledgments}
This work is supported by National Natural Science Foundation of China (Grant No. 11774067 and 11934002), National Program on Key Basic Research Project of China (Grant No. 2021YFA1400900), Shanghai Municipal Science and Technology Major Project (Grant No. 2019SHZDZX01), Shanghai Science Foundation (Grant No. 19ZR1471500).

\end{acknowledgments}

\bibliography{references}

\end{document}